\documentstyle[12pt,aasms4,flushrt]{article}
\begin{document}

\def\plottwox#1#2{\centering \leavevmode
\epsfxsize=.5\columnwidth \epsfbox{#1}
\epsfxsize=.5\columnwidth \epsfbox{#2}}

\def\kms{\rm km\,s^{-1}}
\def\erg{\rm erg}
\def\ergs{\rm erg\,s^{-1}}
\def\Lx{L_{\rm X}}
\def\Lb{L_{\rm B}}
\def\fb{f_{\rm B}}
\def\fx{f_{\rm X}}
\def\NH{N_{\rm H}}
\def \eg           {{e.g.}}
\def \date         {\ifcase\month \message{zero} \or
                    January \or February \or March \or April \or May \or June 
                    \or July \or 
                    August \or September \or October \or November \or 
                    December \fi
                    \space\number\day, \number\year}

\def\Halpha{H$\alpha$}
\def\Hbeta{H$\beta$}
\def\Hgamma{H$\gamma$}
\def\OI{[O\thinspace{\sc i}]\thinspace$\lambda$6300}
\def\OII{[O\thinspace{\sc ii}]\thinspace$\lambda$3727}
\def\OIIIone{[O\thinspace{\sc iii}]\thinspace$\lambda$4959}
\def\OIIItwo{[O\,{\sc iii}]\thinspace$\lambda$5007}
\def\OIIIthree{[O\thinspace{\sc iii}]\thinspace$\lambda$4363}
\def\OIIIth{\OIIIthree}
\def\OIII{[O\thinspace{\sc iii}]\thinspace$\lambda\lambda$4959,\thinspace5007}
\def\NeIII{[Ne\thinspace{\sc iii}]\thinspace$\lambda$3869}
\def\NII{[N\thinspace{\sc ii}]\thinspace$\lambda$6584}
\def\SIIone{[S\thinspace{\sc ii}]\thinspace$\lambda$6716}
\def\SIItwo{[S\thinspace{\sc ii}]\thinspace$\lambda$6731}
\def\SII{[S\thinspace{\sc ii}]\thinspace$\lambda$6716,6731}
\
\def\remark#1{{\bf (#1)}}

\title{THE NATURE OF THE DIFFUSE CLUMPS AND THE X-RAY COMPANION 
OF MRK~273$^{1}$}
\author{X.-Y. Xia$^{2,9}$, S. Mao$^{3,9}$, H. Wu$^{4,9}$,
X.-W. Liu$^{5,9}$, Y. Gao$^{6,7}$, Z.-G. Deng$^{8,9}$, Z.-L. Zou$^{4,9}$}
\altaffiltext{1} {Based in part on observations made with the William Herschel
	Telescope, which is operated on the island of
	La Palma by the Isaac Newton Group, in
	the Spanish Observatorio del
	Roque de los Muchachos of the Instituto de 
	Instituto de Astrof\'\i sica de Canarias.}
\altaffiltext{2}{Dept. of Physics, Tianjin Normal University,
        300074 Tianjin, China}
\altaffiltext{3}{Max-Planck-Institut f\"ur Astrophysik,
        Karl-Schwarzschild-Strasse 1, 85740 Garching, Germany}
\altaffiltext{4}{Beijing Astronomical Observatory,
        Chinese Academy of Sciences, 100080 Beijing, China}
\altaffiltext{5}{Dept. of Physics and Astronomy, 
University College London, Gower Street, London WC1E 6BT, U.K.}
\altaffiltext{6}{Dept. of Astronomy, University of Illinois, 1002
        W. Green St., Urbana, IL 61801, USA}
\altaffiltext{7}{Dept. of Astronomy,
	60 St. George Street, Toronto, Ontario M5S 3H8, Canada}
\altaffiltext{8}{Dept. of Physics, Graduate School,
        Chinese Academy of Sciences, 100039 Beijing, China}
\altaffiltext{9}{Beijing Astrophysics Center, 100871 Beijing, China}

\received{\date}
\accepted{ }

\begin{abstract}

We present an optical spectrum 
for Mrk~273x, the X-ray source $1\arcmin.3$ 
to the northeast of Mrk~273.
The new spectrum indicates that the object is at a much 
higher redshift (0.458) than the value (0.0376) previously reported by us. 
All the detected emission lines 
show properties of Seyfert 2 galaxies. Mrk\,273x has
one of the highest X-ray and radio luminosities 
($L_{\rm x}\approx 1.1\times 10^{44}\,\ergs$, 
$L_{\rm 1.37 GHz}\approx 2.0\times 10^{40}\,\ergs$) among Seyfert 2 galaxies,
yet it has a low neutral hydrogen column density, 
$N_{\rm H} \approx 4.4\times 10^{20}$\,cm$^{-2}$.
These properties
seem difficult to explain in the unified scheme of active galactic nuclei. 

The spectrum which we previously used to identify the 
redshift for Mrk~273x turns out to be for 
diffuse clumps in the northeast tail/plume about $20''$ ($\sim 20$ kpc
in projected distance) from the nuclear region of Mrk~273.
A new spectrum for this region was obtained;
this spectrum is essentially the same as the previous one.
These observations indicate
that these clumps have the same redshift as Mrk\,273 and 
are physically associated with the merger. 
The spectrum is dominated by strong emission from the [O~{\sc iii}] doublet
with Seyfert 2-like line ratios, which indicate that
these northeast clumps are probably excited by the shock plus
precursor mechanism
during the merging process. This mechanism may be operating in other 
ultraluminous IRAS galaxies as well. 
\end{abstract}

\bigskip

\keywords{galaxies: Seyfert -- galaxies: active -- galaxies:
individual (Mrk~273) -- galaxies: interactions -- galaxies: ISM
-- X-rays: galaxies}

\clearpage

\section{INTRODUCTION}

Mrk~273 is an ultraluminous infrared galaxy with prominent
merging signatures. Fig.~1 shows our scanned POSS II J-band image 
and a deep R-band image (Hibbard \& Yun 1999, kindly made available 
to us by John Hibbard) of the Mrk~273 field.
These optical images reveal a long tail to the south and a faint plume
to the northeast.
The Very Large Array (VLA) continuum observations
revealed extremely extended radio lobes up to more than ~200 kpc
in length extending further along the tip of the southern optical 
tail into southeast (Yun 1997;
Yun \& Hibbard 1999). VLA H\thinspace{\sc i} 21\,cm line 
observations show extended atomic gas associated with the southern 
optical tail as well (Hibbard \& Yun 1996, 1999).
Mrk~273 is also detected in the X-ray by ROSAT and
ASCA (Turner et al. 1993, 1997, 1998; Iwasawa 1998). ROSAT HRI image reveals
an X-ray companion source of comparable brightness as
Mrk~273, about $1\arcmin.3$ to the northeast of
Mrk~273. Xia et al. (1998) named this source Mrk~273x, and found an
optically faint source coincident with the X-ray position 
(cf. Fig.~1 in Xia et al. 1998; the position is marked here 
by `X' in Fig.~1a and is very prominently shown 
in the deep R-band image in Fig.~1b). All these optical features are also 
fairly prominent in the Digitized Sky Survey (DSS) image of Mrk~273.

The core of Mrk~273 has been resolved into two compact components
in the radio and near-infrared (Condon et al. 1991; Majewski et al. 1993; 
Knapen et al. 1997), which are believed to be the
double nuclei of the two progenitor galaxies involved in the merging 
and at least one of them is a Seyfert 2 nucleus (Sargent 1972; 
Asatrian et al. 1990).
Spectroscopic observations were carried out on the 2.16m telescope
at the Beijing Astronomical Observatory (BAO) to secure the redshift 
of Mrk~273x and the BAO spectrum indicates that an object 
(thought to be Mrk~273x) at $z=0.0378$ was detected (Xia et al. 1998).
Since the signal-to-noise ratio of this
spectrum was low, further observations of Mrk~273x were performed 
at the William Herschel Telescope (WHT) for additional check. 
The new spectrum turned out to be
different from the one obtained at BAO. Later examinations and
further spectroscopic observations show
that the object detected in the initial BAO spectrum was not Mrk~273x,
but an object about $20$\arcsec\, to the northeast of Mrk~273; 
we will call this diffuse object Mrk~273D. The mis-identification 
was caused by incorrect slit positioning in previous observation 
at BAO (see Section 2).

In this paper, we present the WHT spectrum of Mrk\,273x and
new observations for Mrk\,273D performed
at BAO. In addition, we re-examine all the original data analyzed by Xia et al.
(1998). We find that both Mrk\,273x and Mrk\,273D have
intriguing properties. We show that the emission line
widths and line ratios of Mrk~273x resemble Seyfert 2
galaxies. Since Mrk~273x turns out to be a distant background source, its 
X-ray and radio luminosities
are one of the highest among Seyfert 2 galaxies and 
yet it has a very low neutral hydrogen column density.
Mrk~273D consists of diffuse clumps at the same redshift as the
ultraluminous galaxy Mrk~273 and has quite unusual
line ratios of \OIIItwo/\Hbeta\, and \NII/\Halpha. These ratios are not
easily produced by photoionization of O, B stars. We argue that
the emission line ratios are better explained by shock excitation with 
precursors (Dopita \& Sutherland 1995).
The structure of the paper is as follows. In \S 2, we describe the WHT
observation for Mrk~273x and new observations for Mrk~273D at BAO. We
also explain the original observational setup at
BAO. In \S 3, we present the results of our spectral analysis.
In \S 4, we discuss the nature of Mrk~273x and Mrk~273D. And finally,
a brief summary of the main results is given in \S 5.
Throughout this paper, we use a Hubble constant of $H_0=50\,\kms {\rm
Mpc}^{-1}$ and an Einstein-de Sitter ($\Omega_0=1$) cosmology.

\section{OBSERVATIONS AND DATA REDUCTION}

The observations carried out at BAO used a Zeiss
universal spectrograph mounted on the 2.16m telescope of BAO. 
A Tektronix 1024x1024 CCD was used as the detector. 
The original slit setup, for the BAO observation on April 12, 1997, 
runs from the northeast to southwest passing through star S and 
source D, is illustrated in Fig.~1a. The
slit was approximately $4\arcmin$ long and $4\arcsec$ wide.
This observational setup was intended to determine
the redshift for Mrk~273x (source X in Fig.~1a).
Unfortunately the slit missed the target due to a slight
error in the slit orientation. This occurred because
Mrk~273x was too faint to be seen on the TV monitor, so
the slit had to be rotated from an initial orientation
(which passed through the star `S' in Fig.~1a) by a prescribed angle
such that it would pass through Mrk\,273x. The amount 
of rotation required was calculated from several bright objects
within the field. This technique of slit positioning
for faint sources had been successfully employed previously at BAO,
so no extra check was enforced in the initial observations reported
in Xia et al. (1998). In spite of the
incorrect rotation angle used, by coincidence, an
object other than star S was detected in the spectrum
and was (incorrectly) identified as Mrk\,273x.
Follow-up observations of Mrk\,273x 
carried out at the William Herschel Telescope (WHT) at La Palma
however yielded spectra that are completely different from that
obtained at BAO (see below).

A re-examination of the long-slit spectrum obtained at BAO showed that
the object detected in the spectrum was separated from the star S
by 80$\arcsec$ rather than 20$\arcsec$, the actual angular
distance between Mrk~273x and the star S. From the angular
distance between the object detected in the BAO spectrum
and the star S, and from the fact that only two objects were seen in the
spectrum, we were able to reconstruct the slit orientation as shown in
Fig.~1a. This slit orientation is well-constrained: a little more
rotation to the southwest (anti-clockwise) the slit would enclose the
bright southern tail of Mrk~273, while slightly more rotation to the
opposition direction it would enclose no object other than the
star S. It turned out that the slit passed through some
bright clumps (source D in Fig.~1a),
about $20\arcsec$ to the nuclear region of Mrk~273 (Knapen et al. 1997).
While this slit re-construction seems convincing, 
as a double check, Mrk\,273D was re-observed at
BAO on February 20, 1999 with a different slit setup. The new 
observation yields a spectrum
nearly identical to the one published by Xia et al. (1998).
The slit setup was recorded with photographs, which
allowed us to determine the slit orientation reliably. This
new slit is approximately $4\arcmin$ long and $2.5\arcsec$
wide and runs nearly vertically (in the north--south direction passing
through stars S1 and S2) as shown in Fig.~1a. 

Both the previous and the new BAO spectra cover a wavelength range
of 3500\AA\ to 8100\AA\ with a grating of 195\AA/mm and
a spectral resolution of 9.3\AA\ (2 pixels). 
Wavelength calibration was carried out using an Fe-He-Ar lamp and
standard stars were observed to perform 
flux calibrations. The wavelength calibration
accuracy is better than 1\AA. 

Mrk~273x (source X in Fig. 1) was first observed on June 19, 1998
with the WHT. Since the resulting spectrum was different
from the one obtained earlier at BAO, another observation was attempted
at the WHT on June 26, 1998 which yielded spectrum identical to the one 
obtained a week ago on June 19. Additional new spectrum for Mrk~273x 
was also obtained using the BAO 2.16m telescope,
confirming the results from the WHT. The new BAO spectrum for Mrk~273x
is not presented here given its lower signal-to-noise ratio. The
WHT spectra were obtained using the double spectrograph ISIS.
The CCD detector used at the Blue and Red Arm was respectively an
EEV chip of 2148$\times$4200 pixels with a 13.5$\mu$m pixel size and
a Tek chip of 1124$\times$1124 pixels with a 24$\mu$m pixel size.
At the Blue Arm, a grating of 64{\AA}/mm centered at 4583{\AA}
yielded a useful wavelength coverage from 3200--5300\AA. The violet end
of the blue spectra was cut off by the atmosphere and by instrumental
optics, while the red end was blocked out by the dichroic which had
a half-power crossover wavelength of 5300{\AA}. For the Red Arm,
a grating of 158{\AA}/mm dispersion covered the wavelength range
5300--8270{\AA}. Three 2400s narrow-slit (0.8\arcsec) exposures and one 1800s 
wide-slit spectra (6\arcsec) were obtained 
for each Arm. The
full width at half maximum (FWHM) spectral
resolution was 6.2{\AA} for the Red Arm and 3.28{\AA} for the Blue
Arm, respectively. Wavelength calibration was
carried out using an Ar-Ne lamp; the resulting wavelength accuracy is
about 0.2\AA. The spectra were flux-calibrated using observations of 
the standard star BD\,+33$^{\rm o}$\,2642. 

All the optical spectral data reduction was performed at BAO
using IRAF packages.
The CCD data reduction includes standard procedures such as
bias subtraction, flat fielding and cosmic ray removal.
The measurements of emission lines were performed under
the IRAF environment using tasks ``splot'' and ``ngaussfit''. 

\section{RESULTS}

\subsection{The Optical and X-ray Properties of Mrk\thinspace273x}

We first show the WHT optical spectrum for Mrk~273x in Fig.~2. The spectrum
combines both the Red and Blue Arm data.  \OIII, \Hbeta, \OIIIthree,
\Hgamma, \NeIII, \OII\ and Mg\,{\sc ii} $\lambda\lambda$2796, 2803
emission lines are all convincingly detected.
The redshift determined from these emission lines is 0.458.
The measured fluxes for the identified lines are listed 
in Table~1. All detected lines have a FWHM
of about $600\,\kms$. The optical
B and R magnitudes for Mrk~273x are respectively about 20.8 and 19.6
from the USNO-A1.0 catalog (Monet 1996). The absolute magnitude
of Mrk~273x is therefore $M_B=-21.6$, comparable to 
$L^\star$ in the Schechter luminosity function of galaxies (e.g., Lin
et al. 1996). This luminosity includes the contribution from both 
the central active galactic nucleus (AGN, see below) and the host 
galaxy. In the short {\it Hubble Space Telescope (HST)} Wide Field
Planetary Camera 2 (WFPC2) snapshot image, the underlying galaxy 
is visible but its outer faint contours are not well-detected.  From 
the deep R-band image (shown here in Fig.~1b)
of Hibbard \& Yun (1999), the AGN nucleus and the host galaxy
seem to contribute comparable amount of light.

As reported in Xia et al. (1998), Mrk~273x is also luminous in 
the soft X-ray (0.1--2.4 keV) band. The soft X-ray flux is
$\fx = 1.1 \times 10^{-13}\,\ergs {\rm cm}^{-2}$, corresponding to
an X-ray luminosity of $\Lx \approx 1.1 \times 10^{44}\,\ergs$.
The soft X-ray spectrum of Mrk~273x is well-fitted by a power-law,
with a photon index of $-1.98$, a value typical for Seyfert 2 
galaxies. Spectral fitting yielded a neutral hydrogen column density of
$N_{\rm H}=(4.3\pm 2.2) \times 10^{20}\,{\rm cm}^{-2}$.

Mrk~273x is also detected at the VLA 21 cm continuum observations
and appears to be consistent with a point-like source
(Yun 1997; Yun \& Hibbard 1999). 
At this high redshift, Mrk~273x falls into the class of powerful 
radio sources with $L_{\rm 1.37 GHz} \approx 2.0 \times 10^{40} \ergs$.
This single property sets Mrk~273x into categories of AGNs, QSOs,
and starbursts and radio galaxies rather than normal galaxies.

The high X-ray luminosity, strong radio emission and the emission 
line widths (FWHM $\approx 600 \kms$) of Mrk~273x imply that the 
main energy output mechanism is AGN, and Mrk~273x may be either 
a narrow-line Seyfert 1 (NLS1) galaxy (Osterbrock \& Pogge 1985)
or a Seyfert 2 galaxy from the relatively 
narrow emission lines. Below we explore each possibility in turn.

NLS1s are defined as Seyfert galaxies which have FWHM for
the \Hbeta~line in the range of 500-2000 $\kms$ and with
\OIIItwo/\Hbeta$<$3. Mrk~273x does not satisfy the second criterion since   
the ratio of the \OIIItwo/\Hbeta\ is about 6. Furthermore,
Mrk~273x does not show prominent Fe II emission lines, which are strong
in most (but not all) NLS1 galaxies. 
Also, the soft X-ray spectrum is well fitted by power-law with a photon
index $-1.98$ (Xia et al. 1998) and is not as steep as 
most NLS1s. Hence, Mrk~273x does not
fit in the definition of NLS1 galaxies. 

The relatively narrow Balmer and forbidden lines and
the line ratios of \OIIItwo/\Hbeta\ $\sim 6$ are consistent with 
Mrk~273x being a Seyfert 2 galaxy.
Turner et al. (1997, 1998) and Polletta et al. (1996) presented catalogs 
for several tens of Seyfert 2 galaxies. In their catalogs, the 
soft X-ray luminosity for all but a few listed Seyfert 2 galaxies 
are less than $10^{44}\,\ergs$ and the $\NH$ value is larger than 
$10^{21}\,{\rm cm}^{-2}$. 
The combination of a low $\NH$ value together with powerful X-ray emission 
($\Lx\approx 1.1 \times 10^{44} \ergs$) is therefore
rare among Seyfert 2 galaxies. This peculiarity
is also supported by its quite high radio luminosity and 
multi-wavelength observations of Mrk~273x. 
The flux ratios of Mrk273x in the soft X-ray, B-band,
and radio are $\fx:\fb:f_{\rm 1.37GHZ}=7:1:1.3 \times 10^{-3}$;
such a spectral energy distribution is rarely seen in any other AGNs
(see Xia et al. 1998 for more discussions). Furthermore, the low $\NH$ value 
is not expected in the unified scheme of AGNs (e.g., Dopita 1997 and
references therein). In this picture,
the broad-line regions of Seyfert 2 galaxies are postulated to be
obscured by a thick torus of gas and dust, which presumably gives rise to
high $\NH$ values.
 
To summarize, the optical emission line properties, the powerful X-ray 
and radio emission and the low neutral hydrogen absorption indicate that 
Mrk~273x is a rare source that may
provide a test of the unified picture of AGNs.

\subsection{Diffuse Clumps In the Tidal Plume of Mrk~273}

The diffuse clumps (Mrk~273D, labeled as source D in Fig.~1a) are
clearly seen in the J-film copy of POSS II and in the DSS image, but 
they are visible neither in the near-infrared (e.g., Smith et al. 1996),
most R-band images (e.g., Yun \& Scoville 1995; Mazzarella \& Boroson 1993),
nor in the HST WFPC2 F814 snapshot image,
suggesting that they have quite blue colors. Only the deep
R-band image shown in Fig.~1b (Hibbard \& Yun 1999) reveals
some patches in the northeast plume corresponding to these dense knots.
We extracted the BAO spectra for the diffuse clumps in Fig.~1a 
using an aperture window 
size of 4\arcsec$\times$ 17\arcsec\, for the observation on April 12,
1997 and 2.5\arcsec$\times$ 17\arcsec\, for the observation
on February 20, 1999. These aperture windows are indicated as
the rectangles in Fig.~1a.
The spectra are shown in Fig.~3 and the emission line fluxes
are listed in Table~1. The old and new spectra are similar. The
continuum is somewhat higher in the Feb. 20, 1999 observation. 
The difference may be real, as the two slits sampled
slightly different regions, although the possibility of
calibration uncertainties cannot be completely ruled out.
The emission line ratios are very similar from both spectra.
The redshift for Mrk~273D determined from the emission lines
is the same as Mrk~273 ($z=0.0376$). Therefore
these clumps are physically associated with the major merger process
of Mrk~273. Yun \& Scoville (1995) suggest
that Mrk~273 is the merging product of
a nearly edge-on gas-rich spiral and another more face-on spiral.
The tidal plume in the northeast is from the face-on 
progenitor. So the physical association of Mrk 273d with Mrk 273 is
expected.

It is clear from Fig.~3 and Table~1 that 
the continuum emission from Mrk~273D is very weak and the \OIII\ 
lines are strong compared to \Hbeta. More specifically, from the 
observation on April 12, 1997,
the line ratio of the \OIIItwo\ to \Hbeta\, is $18.6^{+9.5}_{-4.9}$,
while the line ratio of the \NII\ to \Halpha\ is $0.34\pm 0.1$
(the error bars are 1$\sigma$ values). For the new observation,
the ratio of the \OIIItwo\ line to \Hbeta\, is $18.9^{+13}_{-5.7}$
while the ratio of \NII\ line to \Halpha\ is  $0.3\pm 0.1$.
Note that these two line ratios
are little affected by dust extinction since the two lines involved 
in the ratios are quite close in wavelength.
Fig.~4 shows the standard diagnistic diagram of \OIIItwo/\Hbeta\ versus
\NII/\Halpha\ (Osterbrock 1989) for emission line galaxies.
The HII, LINER and Seyfert galaxies occupy different regions in this diagram.
For HII regions photoionized by O, B stars,
$[{\rm O}\,{\sc iii}]\thinspace\lambda5007/{\rm H}\beta < 5$ for
$[{\rm N}\thinspace{\sc ii}]\thinspace\lambda6584/{\rm H}\alpha
\approx 0.3-0.4$ (see also Fig.~12.1 in Osterbrock 1989). The
\OIIItwo/\Hbeta\, ratio of Mrk 273D is clearly much higher 
than those typically found for photoionized H~{\sc ii} regions.
In fact, Mrk 273D is clearly located in the region occupied by
Seyfert 2 galaxies. The \OI/\Halpha\ and \SII/\Halpha\, line ratios of
Mrk 273D are also located in the Seyfert 2 region (cf. Fig.~12.2,
12.3, Osterbrock 1989 and more discussions in Xia et al. 1998).
To achieve these Seyfert 2 like line ratios,
the ionization source must be harder than the radiation provided 
by young massive stars.
Mrk~273D is, however, obviously not an AGN
given its diffuse morphology.
We discuss further the mechanism of line excitations
in Mrk~273D in \S 4.2.

\section{DISCUSSION}

\subsection{Mrk~273x: an Unusual Seyfert 2 Galaxy}

Mrk~273x is a background object in the Mrk~273 field; it is at a much
higher redshift ($z=0.458$) than Mrk~273 which has a redshift of
0.0376. It is interesting to note that
the X-ray companions of three nearest ultraluminous IRAS galaxies
(Arp~220, Mrk~273 and Mrk~231) are all background sources and
are therefore not physically associated with the mergers themselves
(cf. Xia et al. 1998). Mrk~273x has AGN characteristics both optically
and in the soft X-ray together with a high radio luminosity. 
The narrow emission
lines and various line ratios are consistent with it being a Seyfert
2 galaxy. Its soft X-ray luminosity is one of the highest among
Seyfert 2 galaxies with a low neutral hydrogen column density. 
Although many well known nearby Seyferts (including Mrk~273 itself)
also host an energetically significant starburst, the low neutral hydrogen
column density of $N_{\rm H} \approx 4.4\times 10^{20}\,{\rm cm}^{-2}$
appears to exclude the possibility of Mrk~273x being a dusty luminous
starburst galaxy.
These observational facts show that Mrk~273x is an unusual Seyfert 2
galaxy that is not easy to explain using the unified scheme of AGNs.

Xia et al. (1998) examined the time variability of Mrk~273x
using the ROSAT data taken in May and June 1992. While Mrk~273x is fainter
than Mrk~273 by about 20\% in the ROSAT PSPC image, it is almost as
bright as Mrk~273 in the ROSAT HRI image. A $\chi^2$ test however
reveals that these changes are not statistically significant
in the ROSAT data. From
the 0.5-2 keV image of SIS on board ASCA, Iwasawa (1998) found that
Mrk~273x is at most 40 percent as bright as Mrk~273 on
Dec. 27, 1994. Since Mrk~273 is not known to vary
in the soft X-ray, these multi-epoch observations imply that
Mrk~273x has faded by a factor of 2 in two and a half years.
Cycle 1 AXAF observations are planned to obtain a high-resolution map of
Mrk~273x and Mrk~273. These
observations will provide further insights on their
X-ray variability, higher energy behaviors and 
spatial distribution of the X-ray emissions.

\subsection{The Excitation Mechanism of Mrk~273D}

The diffuse clumps in Mrk~273D in the northeast plume could directly come from
the extreme outer regions of the face-on progenitor.
It is also possible that Mrk~273D was formed in the major merger process. 
This patchy object is unlikely to
be a self-gravitating (tidal) dwarf galaxy since
they are usually found far from the interacting
parent galaxies and their spectra resemble typical photoionized
H~{\sc ii} regions (e.g., Duc \& Mirabel 1997). Furthermore, clumps 
formed close to the merger is liable to tidal disruption
once they fall back to the merger. (However, some observations 
indicate that clumps close to the merging pair do exist, e.g.,
H~{\sc ii} regions in the NGC~4676A tail, Hibbard \& van Gorkom 1996.)
In contrast, Mrk~273D is very close to ($\sim 20$ kpc in projected
distance from) 
the main merging nuclei, and its spectrum differs significantly 
from photoionized H~{\sc ii} regions.

Although the spectrum of Mrk~273D resembles that of a Seyfert 2
galaxy,  its diffuse morphology suggests that it is not a (dwarf) AGN.
In fact, the spectrum of Mrk~273D is very similar
to that of the soft X-ray nebula to the north of Mrk~266 (cf. Fig.~5
in Kollatschny \& Kowatsch 1998). 
As pointed out by these authors, the $V-R$ color
($0.5\pm 0.1$) of the northern component is exceptionally blue due to the
intense \OIII\ line emission.  The diffuse gas
clumps in Mrk~273D are also very blue, since
they are only seen in the B-band, but not in other bands redder than R.
The blue color of Mrk~273D is also due to the continuum being 
dominated by the \OIII\ line emission (cf. Fig.~3).
The spectrum of Mrk~273D is also similar to
the average spectrum of the northwest cone of
NGC~2992 (cf. Fig.~3a in Allen et al. 1999)
and the spectrum of diffuse ionized gas in NGC~891 (Rand 1997, 1998).
As discussed by these authors, the likely excitation mechanism
for these peculiar spectra is the shock plus precursor model
(Dopita \& Sutherland 1995). For a fast shock
with velocity of several hundred $\kms$, copious 
UV photons are produced in the shock front, which can in turn excite
the gas of H~{\sc ii} regions in front of the shock,
thus producing radiative precursors. The hardness of the UV radiation
(shock temperature) depends on the velocity of the shock wave,
and with suitable parameters, 
this scenario can produce the line ratios as seen in Mrk~273D (cf.
Fig.~2a in Dopita \& Sutherland 1995).

It is believed that starbursts can drive radial
outflows (Wang, Heckman \& Weaver 1997).
There is some evidence that such outflows (superwinds) indeed exist
in several ultraluminous IRAS galaxies
out to tens of kpc with velocities of a few hundred 
to 1000 $\kms$, e.g., in NGC~6240, Arp~299, Arp~220
(Heckman et al. 1987, 1990, 1996, 1999;
Schulz et al. 1997; Wang et al. 1997). Since 
Mrk~273 is also a major merger ultraluminous infrared galaxy and has extended 
soft X-ray emission and H\,{\sc i} nebula, it is conceivable that such
an outflow exists in Mrk~273. In this regard,
Mrk~266 is also a candidate since it is a luminous infrared
merging galaxy with double nuclei ($\sim 7$ kpc in projected separation) and
very extended (about 150 kpc) soft X-ray nebula.

These observations suggest 
that the shock+precursor emission may be a common mechanism 
to excite gaseous nebulae in luminous infrared merging galaxies,
in addition to the O, B star photoionization and AGN excitation.
Wu et al. (1998) showed that many observed 
ultraluminous IRAS galaxies have
mixture types in different line ratio diagnostic diagrams.
Goncalves et al. (1998) also suggested that most emission line galaxies with 
the so-called transition spectrum
have composite spectra with simultaneous presence of Seyfert, LINER
and H~{\sc ii} region contributions. Perhaps
these objects are not only just a mixture of AGN and 
photoionized starburst regions (Genzel et al. 1998), they may
also contain shock+precursor regions, as
seen in Mrk~273D and the northern nebula of Mrk~266.
This also provides a caution: a Seyfert 2 like spectrum
in high-redshift luminous infrared galaxies does not
necessarily mean the presence of AGNs at their centers, 
instead such a spectrum could be induced by the shock+precursor excitation
in the gas clumps in the mergers.

\section{SUMMARY}

We have presented optical spectroscopic observations for the X-ray companion
source and the blue diffuse clumps in the northeast 
tidal debris surrounding the ultraluminous galaxy Mrk~273. 
Their peculiar properties are 
discussed and explored together with soft X-ray and radio 
observations available. We summarize below the main 
points we have presented and addressed in this paper.

Mrk~273x, the X-ray companion $1\arcmin.3$ to the northeast of the
ultraluminous galaxy Mrk~273, is a background 
source at $z=0.458$. Its redshift
is much higher than the redshift ($z=0.0376$) of Mrk~273.
The soft X-ray spectrum for Mrk~273x is typical of Seyfert 2 galaxies. 
The X-ray luminosity of Mrk~273x is exceptionally high for
a Seyfert 2 galaxy, $L_{\rm x}\approx 1.1\times 10^{44}\,\ergs$. Spectral
fitting gives a low neutral hydrogen column density, 
$N_{\rm H} \approx 4.4\times 10^{20}$\,cm$^{-2}$.
All the optical emission lines detected for Mrk~273x have a similar FWHM
of $\approx 600$\,km\,s$^{-1}$ and the
\OIIItwo/\Hbeta\, ratio is about 6, again typical for Seyfert 2 galaxies.
Mrk~273x is also a powerful radio source with a radio luminosity 
$L_{\rm 1.37 GHz}\approx 2.0\times 10^{40}\,\ergs$. This adds more
peculiarities to this energetically unusual Seyfert 2 galaxy.

The optical knots Mrk~273D in the northeast tidal tail/plume $\sim 20$ kpc
(projected) from the nuclear region of Mrk~273 is at the same redshift
as that of Mrk~273. They are diffuse gas clumps physically associated
with the major merger. The spectrum of Mrk~237D is
dominated by strong emission from the [O~{\sc iii}] doublet.
The strong line emission gives rise to
the blue color of this object. The [O~{\sc iii}]$\lambda$5007/H$\beta$ and
[N~{\sc ii}]$\lambda$6584/H$\alpha$ line ratios are $\approx 19$ and
0.3, respectively. These Seyfert 2--like line ratios for Mrk~273D 
are likely excited by
shocks plus precursor mechanism involved in the merging process.

\acknowledgments

We are grateful to Zheng Zheng for assistance in observations and preliminary
data reductions. We also thank Simon White for advice and
the anonymous referee and the scientific editor John Huchra for 
valuable criticisms that have improved the paper. We appreciate
very much for the deep R-band image kindly provided to us by John Hibbard.
This project was partially supported by the NSF of China and 
the exchange program between NSFC and DFG. Y.G.'s 
research at LAI, Dept. of Astronomy is funded by NSF grants
AST96-13999 and by the University of Illinois. Y.G. is also grateful 
to Ernie Seaquist for support at the University of Toronto.

\clearpage

\section*{FIGURE CAPTIONS}

\figcaption[fig1.ps]{
(a) (Left) Image of the Mrk~273 field scanned from a J-film copy of 
POSS II. North is up and east is to the left.
The slit for the April 12, 1997 observation
is $4\arcmin$ long and runs from the top left to the bottom
right. The slit direction for the February 20, 1999 observation
runs vertically. The symbols `S', `S1', `S2' are for three reference
stars and `X' is for Mrk~273x. The slit for the
April 1997 observation clearly
missed the intended target, Mrk~273x, while enclosing some diffuse
emissions (labeled as `D', named as Mrk~273D in this paper). 
Spectra are extracted from the central rectangular regions from these two
observations and are shown in Fig. 3. A part of 
the long-slit spectrum for the February 1999 observation
is shown at the bottom panel. Notice that the line emissions dominate
over the continuum. The spectrum for Mrk~273x is shown in Fig. 2.
(b) (Right) A deep R-band image (Hibbard \& Yun 1999, kindly 
provided by J.E. Hibbard) of the same field as Fig.~1a.
Both Mrk~273x and Mrk~273D can be clearly seen.
Mrk~273D is essentially
absent in most published (shallower) images in the R-band and longer
passbands in the literature.
\label{fig1}}

\figcaption[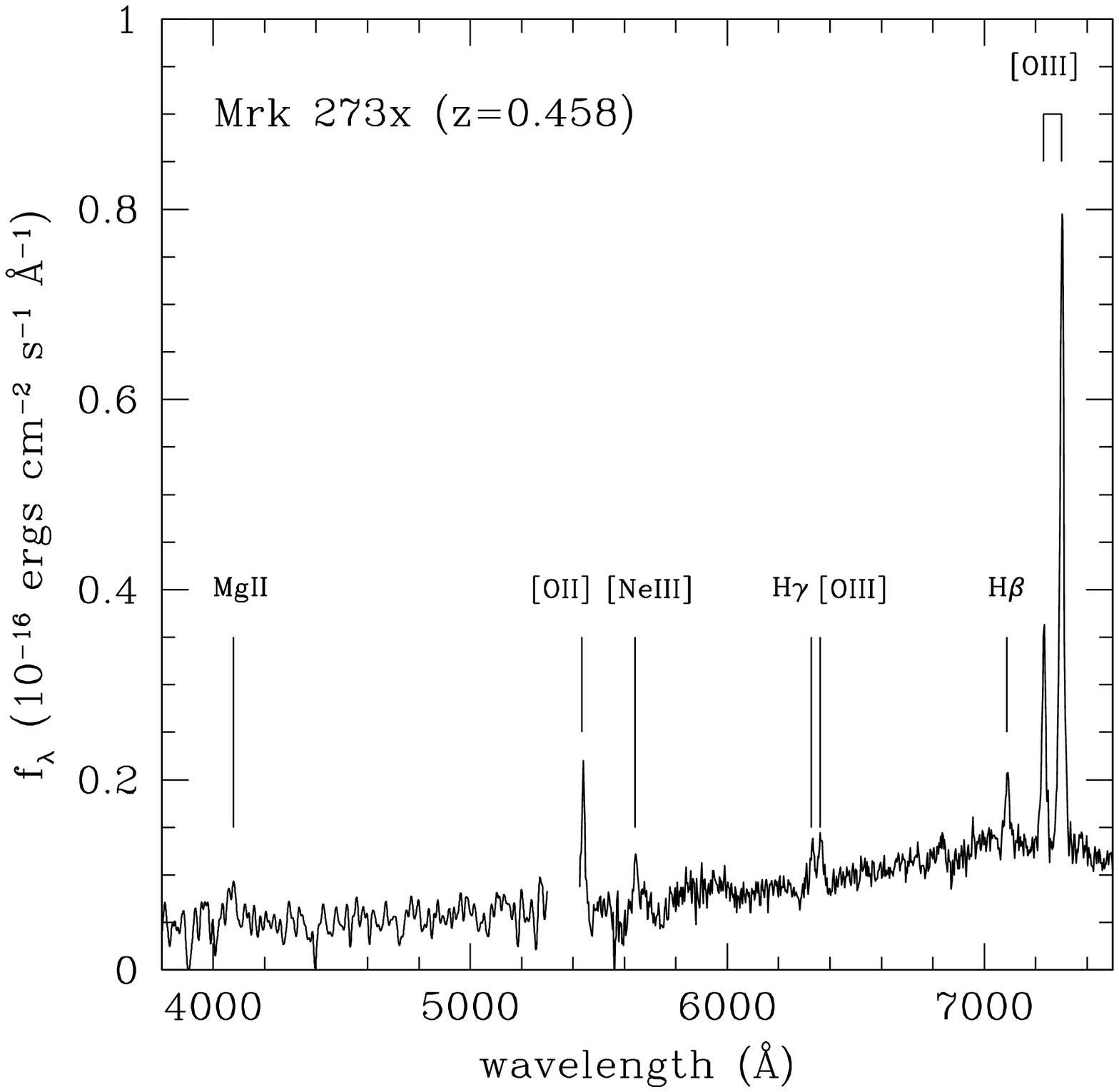]{
Spectrum for Mrk~273x (the object marked as `X' in Fig. 1)
obtained with the William Herschel Telescope in June 1998. Only the high
signal-to-ratio part of the spectrum is shown. The gap in the spectrum
is due to the sensitivity drop off in the Blue Arm close to 5300\AA.
Prominent emission lines are labelled.
\label{fig2}
}

\figcaption[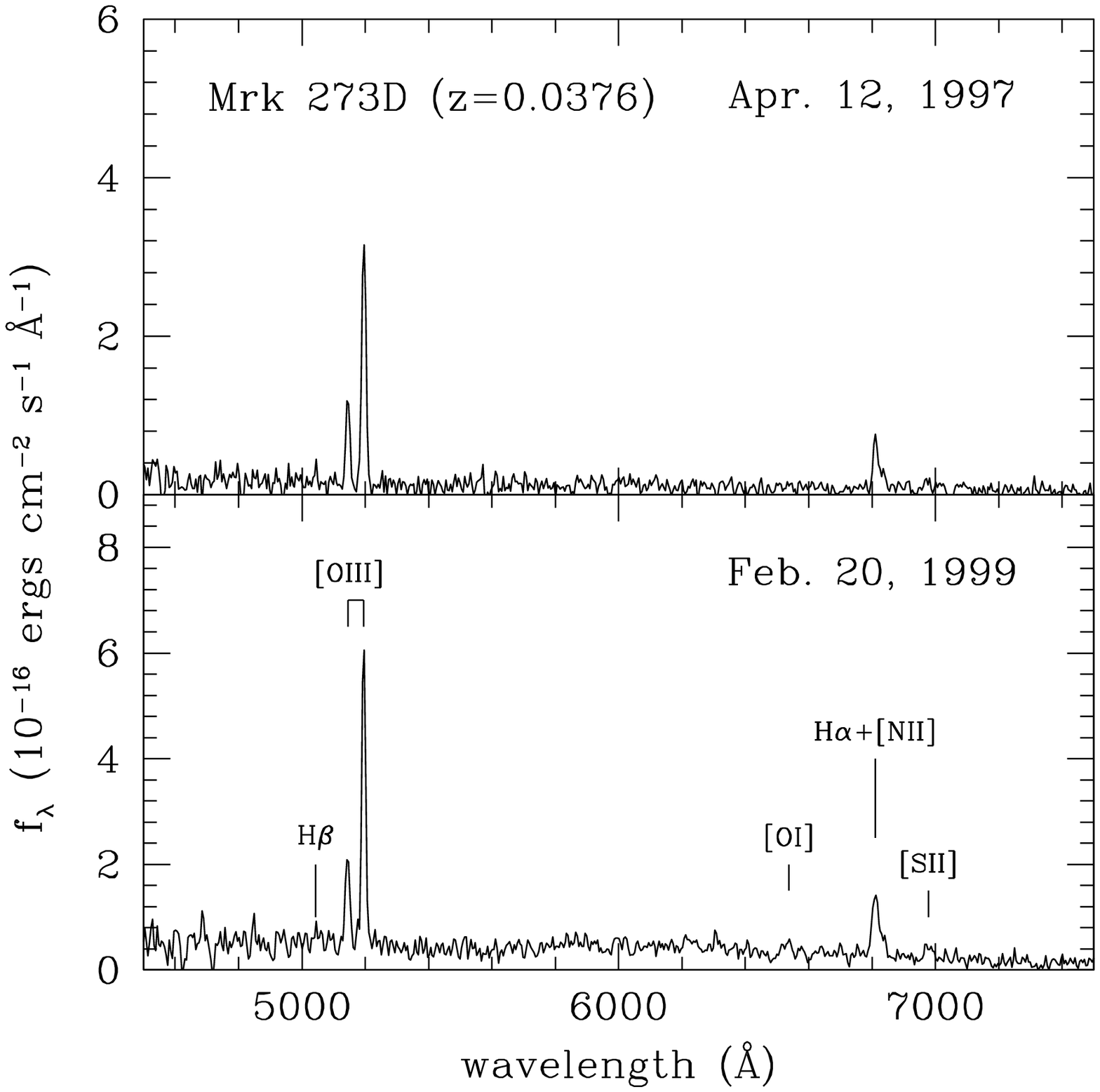]{
Optical spectra for Mrk~273D (cf. Fig. 1).
The top panel shows the  spectrum obtained from the April 12, 1997
observation while the bottom panel shows that obtained from
the February 20, 1999 observation. Emission lines are labelled. 
\label{fig3}
}

\figcaption[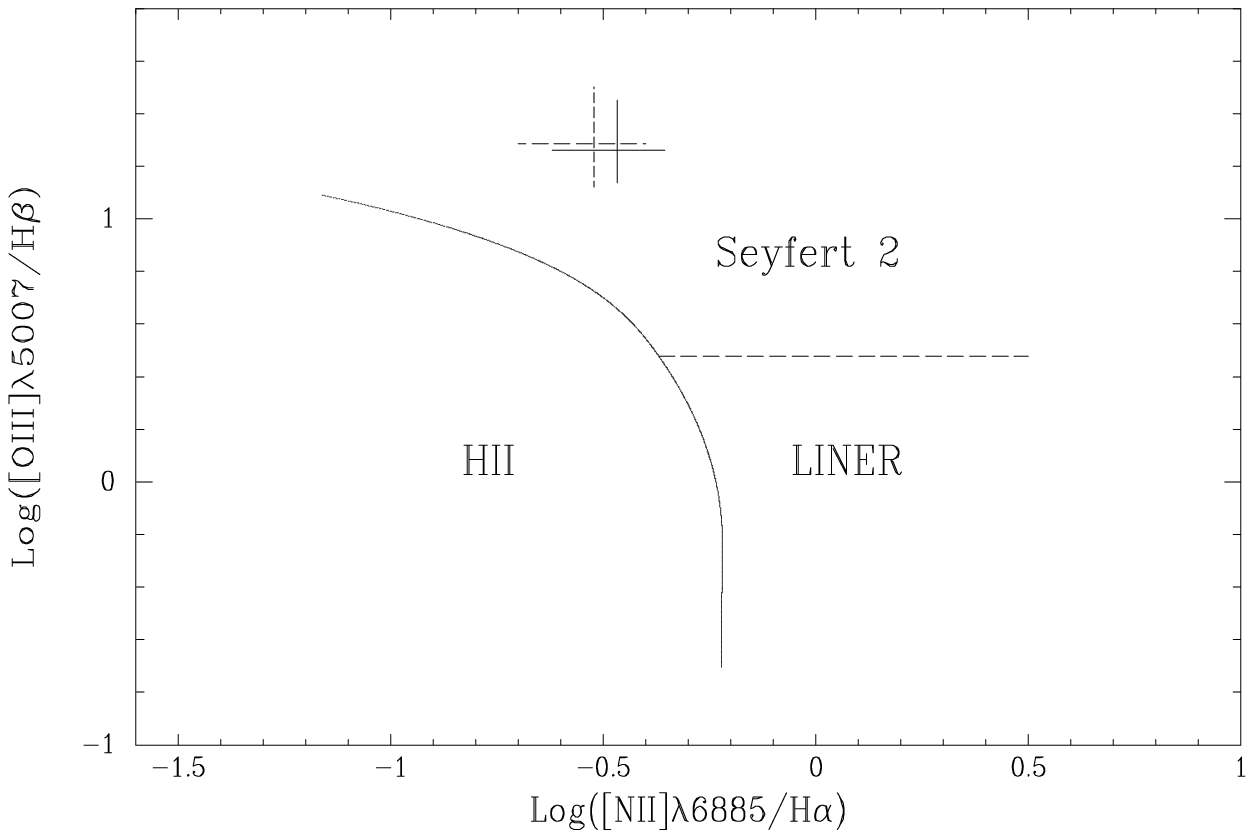]{
\OIIItwo/\Hbeta\ vs. \NII/\Halpha\ diagnostic
diagram used to determine the dominant source
of excitation for emission line galaxies (e.g., Osterbrock 1989).
The horizontal dashed line separates low excitation from high excitation
galaxies, while the solid curve is the empirical dividing line
between HII regions ionized by O, B stars and AGN. 
The two crosses indicate the line ratios obtained in 
the April 1997 (solid) and Februrary 1999 (dashed) observations for Mrk 273D.
\label{fig4}
}

\clearpage

\begin{deluxetable}{lrrr}
\tablecaption{Emission Line fluxes for Mrk~273x and Mrk~273D}
\tablehead{
\colhead{line} & \colhead{Mrk273x} & \colhead{Mrk273D I} &
\colhead{Mrk273D II}}
\startdata
\OIIItwo&$13.8\pm 0.7$	&$52.2\pm 2.0$	&$85\pm 4.0$\nl  
\OIIIone&$4.4\pm 0.2$	&$18.4\pm 2.0$  &$31\pm 3.0$\nl
\OI	&\nodata	&\nodata	&$5.3\pm 2.0$\nl
\Hbeta	&$2.4\pm 0.2$	&$2.8\pm 1.0$   &$4.5\pm 2.0$\nl
\OIIIth	&$1.2\pm 0.2$	&\nodata	&\nodata	\nl
\Hgamma	&$1.0\pm 0.2$	&\nodata	&\nodata	\nl
\NeIII	&$1.2\pm 0.2$	&\nodata	&\nodata	\nl
\OII    &$3.0\pm 0.6$	&$6.0\pm 2.0$	&\nodata	\nl
\Halpha	&\nodata	&$11.9\pm 1.0$	&$27.1\pm 3.0$	\nl
\NII	&\nodata	&$4.0\pm 1.0$	&$8.2\pm 3.0$	\nl
\SIIone	&\nodata	&$1.9\pm 0.5$	&$4.9\pm 1.0$	\nl
\SIItwo	&\nodata	&$1.1\pm 0.5$	&$3.7\pm 1.0$	\nl
\tablecomments{
All fluxes are in units of
$10^{-16}\,{\rm erg ~s^{-1}~cm^{-2}}$. The second and third columns 
are respectively for the April 12, 1997 and February 20, 1999
observations of Mrk~273D.
}
\enddata
\end{deluxetable}

\clearpage
\plottwox{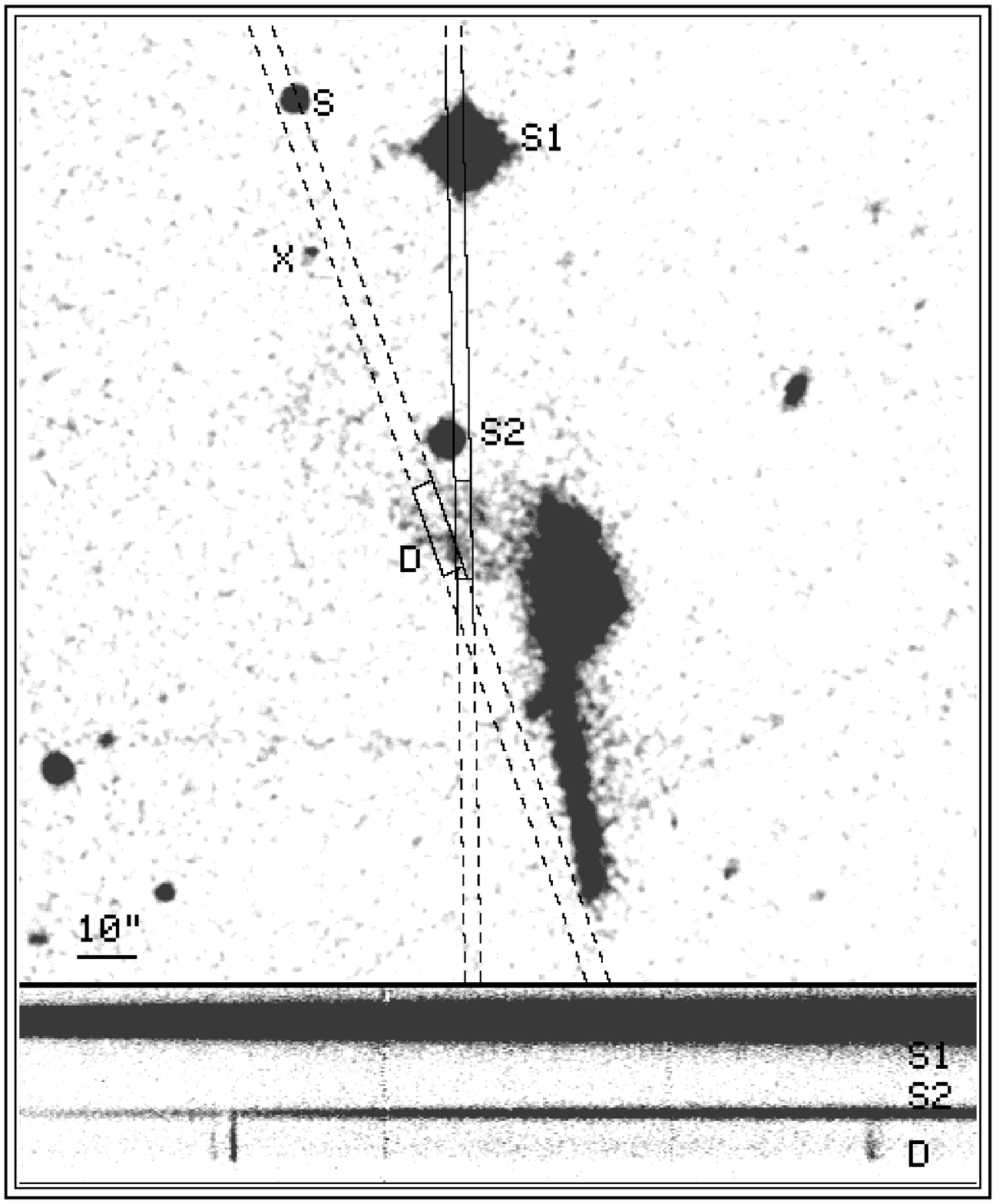}{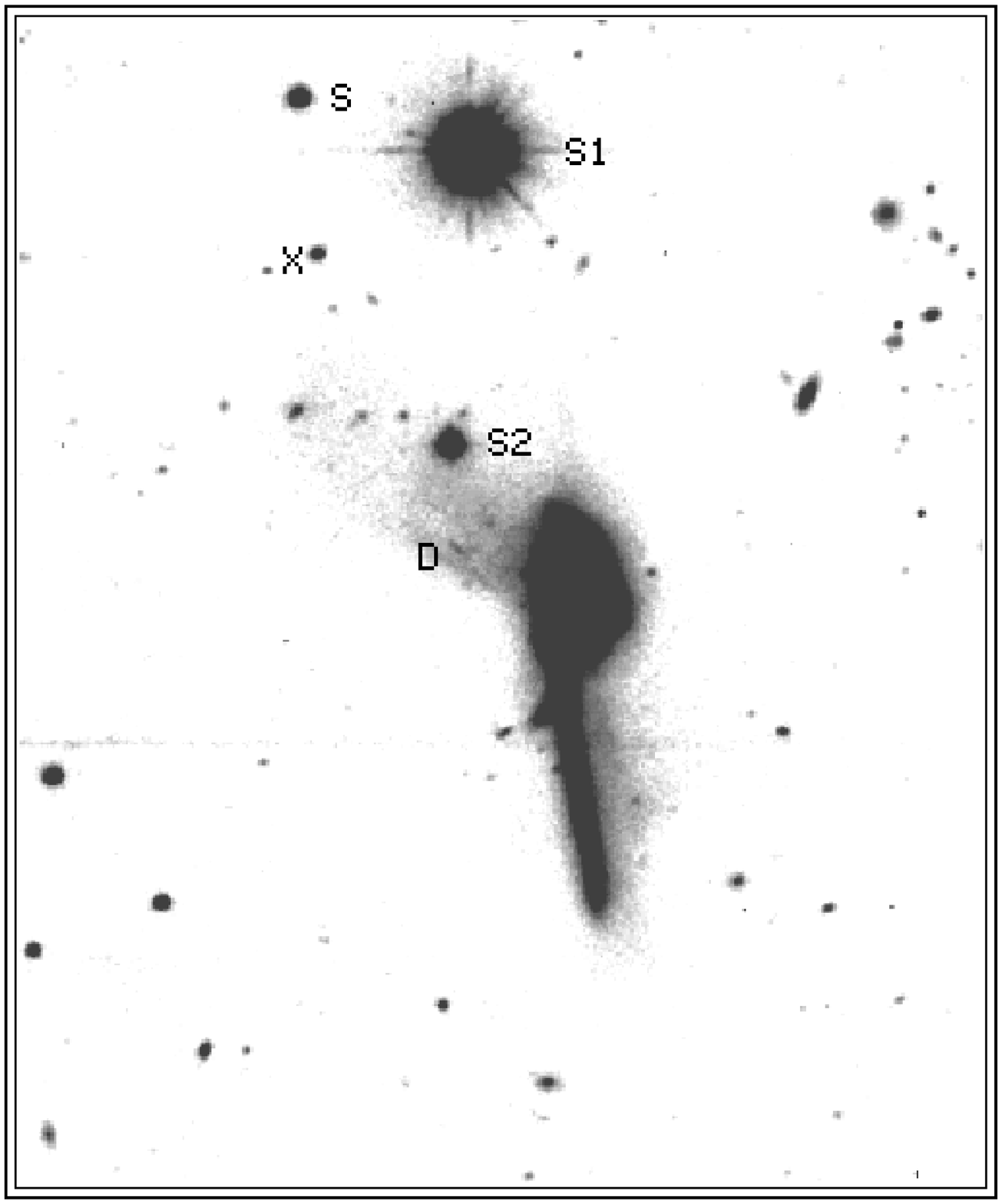}

\clearpage
\plotone{fig2.ps}

\clearpage
\plotone{fig3.ps}

\clearpage
\plotone{fig4.ps}

\end{document}